\documentclass[journal]{IEEEtran}
\IEEEoverridecommandlockouts
\usepackage{lettrine}
\usepackage{cite}
\usepackage{amsmath,amssymb,amsfonts}
\usepackage{algorithm}
\usepackage{algorithmic}
\usepackage{bm}
\usepackage{graphicx}
\usepackage{subfigure}
\usepackage{tabularx}
\usepackage{textcomp}
\usepackage{xcolor}
\usepackage{stfloats}
\usepackage{balance}
\usepackage{multirow}
\def\BibTeX{{\rm B\kern-.05em{\sc i\kern-.025em b}\kern-.08em
		T\kern-.1667em\lower.7ex\hbox{E}\kern-.125emX}}

\floatname{algorithm}{\bf{Algorithm}}

\begin{document}
    \setlength{\abovedisplayskip}{4.5pt plus 2pt minus 1.5pt}
    \setlength{\belowdisplayskip}{4.5pt plus 2pt minus 1.5pt}
    \setlength{\abovedisplayshortskip}{3pt plus 1.5pt minus 1pt}
    \setlength{\belowdisplayshortskip}{3pt plus 1.5pt minus 1pt}
    \title{Rank-Aware Element Grouping for Power-Efficient Multiuser ISAC With an Extremely Large-Scale IRS}
	\author{Shengsheng~Zhang,~\IEEEmembership{Graduate~Student~Member,~IEEE,} Ritao~Cheng, Zitong~Wang, \\~\IEEEmembership{Graduate~Student~Member,~IEEE,} Cheng Zhang,~\IEEEmembership{Member,~IEEE,} Meng~Hua,~\IEEEmembership{Senior~Member,~IEEE,} and~Luxi~Yang,~\IEEEmembership{Senior~Member,~IEEE}%
		\IEEEcompsocitemizethanks{
        \IEEEcompsocthanksitem This work was supported by the National Science and Technology Major Project of China under Grant 2025ZD1302500, the Natural Science Foundation on Frontier Leading Technology Basic Research Project of Jiangsu under Grant BK20222001, and the National Natural Science Foundation of China under Grants U1936201 and 61971128. 
        \IEEEcompsocthanksitem Shengsheng Zhang, Zitong~Wang, Cheng Zhang and Luxi Yang are with the School of Information Science and Engineering, the National Mobile Communications Research Laboratory, and the Frontiers Science Center for Mobile Information Communication and Security, Southeast University, Nanjing 210096, China, and also with the Pervasive Communications Center, Purple Mountain Laboratories, Nanjing 211111, China (e-mail: zhangshengsheng@seu.edu.cn; ztwang2410@seu.edu.cn; zhangcheng\_seu@seu.edu.cn; lxyang@seu.edu.cn). 
        \IEEEcompsocthanksitem Ritao Cheng is with the China Mobile Group Design Institute Co., Ltd., No. 16 Danling Street, Haidian District, Beijing 100080, China (e-mail: chengritao@cmdi.chinamobile.com).
		\IEEEcompsocthanksitem Meng Hua is with the Department of Electrical and Electronic Engineering, Imperial College London, London SW7 2AZ, UK (e-mail: m.hua@imperial.ac.uk).
	}}
	
	\maketitle
	
	\begin{abstract}
        We investigate power-efficient multiuser integrated sensing and communication (ISAC) assisted by an element-grouping extremely large-scale intelligent reflecting surface (EG-XL-IRS). The grouping pattern is designed using slowly varying statistical channel state information (S-CSI), so that both IRS-related channel acquisition and online passive beamforming operate in the group domain rather than the element domain. We reveal a fundamental gain-rank tradeoff induced by element grouping: phase-consistent grouping can coherently enhance selected deterministic propagation components, while excessive concentration on a common deterministic mode can reduce the effective spatial rank of the multiuser channel and, for extended targets, the diversity of desired-scatterer responses. Motivated by this observation, we develop a task-adaptive rank-aware grouping strategy that balances weak-user enhancement and target-scatterer illumination while preserving task-relevant spatial dimensions. For each candidate grouping pattern, the transmit covariances and group-wise reflection phases are jointly optimized under communication and sensing quality-of-service constraints, followed by physical phase recovery and feasibility verification. Numerical results show that the proposed design substantially reduces the required transmit power compared with representative grouping benchmarks under the same grouping dimension and online optimization budget.
    \end{abstract}
	
	\begin{IEEEkeywords}
		Integrated sensing and communication (ISAC), element-grouping extremely large-scale intelligent reflecting surface (EG-XL-IRS), extended target, effective rank, signal-to-clutter-plus-noise ratio (SCNR).
	\end{IEEEkeywords}
	
    \section{Introduction}
    \IEEEPARstart{I}{ntegrated} sensing and communication (ISAC) enables a common wireless platform to support data transmission and environmental sensing by sharing spectrum, radio-frequency hardware, and signal-processing resources \cite{9737357,11223707,10630588}. In multiantenna ISAC systems, performance depends not only on the available waveform degrees of freedom (DoFs) \cite{9124713}, but also on the spatial structure of the propagation channels. Multiuser communication benefits from sufficiently separable effective user channels, whereas sensing requires target responses that are distinguishable from clutter and, for extended targets, sufficiently diverse across scattering centers \cite{Tse_Viswanath_2005,6649991}. Transmit-power minimization in ISAC therefore depends on both transmit-side resource allocation and the propagation structure available to the communication and sensing functions.
    
    Intelligent reflecting surfaces (IRSs) can reconfigure the reflected links among the ISAC base station (ISAC-BS), communication users (CUs), and the sensing scene \cite{8811733,10555049,9998527,10284917}. Existing IRS-assisted ISAC designs commonly optimize element-wise reflection coefficients based on full or near-perfect instantaneous channel state information (CSI) \cite{8937491,9039554,10018442,9729398,9677923,10319318}. Acquiring the associated cascaded CSI generally incurs an overhead that increases with the number of passive reflecting elements, even when time-division-duplex reciprocity is exploited \cite{9090356,9866003}. For extremely large-scale IRSs (XL-IRSs), both channel acquisition and real-time element-wise reflection optimization become high-dimensional \cite{9306896,9839429}. For XL-IRS-assisted ISAC, a lower-dimensional representation is needed to reduce channel-acquisition and online-control dimensions while retaining propagation structure relevant to both communication and sensing.
    
    Element grouping provides such a lower-dimensional representation. Consider an XL-IRS with $N$ passive elements partitioned into $Q$ groups, where $Q\ll N$, with all elements in each group sharing one reflection coefficient \cite{9039554,10018442,11396439}. The online passive-beamforming dimension is then reduced from $N$ to $Q$. Under the group-wise channel-estimation protocol adopted in this work, only the post-grouping cascaded channels are estimated, so the IRS-related training dimension scales with $Q$ rather than $N$. The partition also changes the effective communication and sensing channels because the element-domain responses are aggregated within each group before fast-timescale optimization. A fixed geometric partition mainly reduces the control dimension, whereas a partition designed from statistical CSI (S-CSI) can use slowly varying spatial phase information to determine which element responses are combined over a slow timescale \cite{9906898}. The resulting element associations affect coherent gain, multiuser channel separability, and, for extended targets, the diversity of the desired-scatterer responses.

    Existing studies treat these aspects separately. Channel-estimation methods focus on reducing CSI-acquisition overhead \cite{8937491,9039554,9839429}, while joint active--passive beamforming optimizes the transmission and reflection variables after the relevant CSI has been acquired \cite{9906898,9913311,10440056,10762897,Xu2025CognitiveRISISAC,Bazzi2025LowDynamicRISISAC,Yigit2025HybridSTARRIS,Zhou2026HybridFieldISAC,Yang2025FDARISISAC,Wu2025MovableRISISAC,Gan2025CoverageRISISAC,Liu2025STARRISFullDuplex}. Existing communication-oriented grouping methods mainly optimize received-power or rate-related criteria \cite{10018442,10286054,9729398,9677923,11396439}. These approaches do not directly address how the element partition should be chosen so that the reduced-dimensional representation retains spatial structure relevant to both multiuser communication and sensing. This issue becomes more involved for extended-target sensing. Maximizing an aggregate target response may concentrate most of the sensing energy on one scattering center, whereas algebraic rank alone does not indicate whether the response energy is concentrated in one dominant singular mode. The grouping design should therefore account for coherent gain, communication-channel separability, and, when multiple desired scattering centers are present, target-response diversity.
    
    Motivated by this observation, we treat element grouping as both a dimensionality-reduction mechanism and a slow-timescale propagation-structure design variable. The grouping matrix is constructed from S-CSI, whereas only the $Q$-dimensional post-grouping cascaded CSI is acquired and the group-wise reflection coefficients are optimized on the fast timescale. This two-timescale design does not require realization-wise knowledge of all $N$ element-domain cascaded coefficients when determining the partition. We consider one desired target represented by $T\geq1$ scattering centers in the presence of a separate set of clutter scatterers. For $T=1$, the grouping design seeks useful-link gain while avoiding excessive concentration of the multiuser channel onto a common spatial mode. For $T>1$, it also needs to avoid concentrating the desired-target response on only a small subset of scattering centers or a single dominant response mode. The proposed design therefore combines S-CSI-based gain criteria with task-dependent structural screening, followed by instantaneous active--passive beamforming under the original communication and sensing quality-of-service (QoS) constraints.
    
    The main contributions are summarized as follows.
    \begin{itemize}
        \item We develop a two-timescale element-grouping XL-IRS (EG-XL-IRS) framework in which an S-CSI-based binary partition reduces the passive control dimension from $N$ elements to $Q$ groups. Under the adopted group-wise channel-estimation protocol, the IRS-related channel-acquisition dimension is also reduced to the group domain, while instantaneous post-grouping CSI is used for fast-timescale active--passive beamforming.
    
        \item Under the adopted channel model, we characterize the gain--rank behavior induced by element grouping. Phase-consistent grouping can coherently enhance selected deterministic propagation components while preserving the total average power of the modeled i.i.d.\ isotropic diffuse component. We further show that, when a common rank-one line-of-sight component dominates the residual channel, the effective rank of the multiuser channel, and for $T>1$ that of the desired-scatterer response matrix, approaches one. This behavior motivates rank-aware screening in addition to coherent-gain and algebraic-rank criteria.
    
        \item We develop task-dependent grouping strategies for point and extended targets by accounting for weak-CU and desired-scatterer gains. Both branches screen candidate partitions according to communication-channel separability. For extended targets, the design also incorporates per-scatterer gain floors and response-diversity safeguards, together with an angle Fisher information matrix (FIM)/Cram\'er--Rao bound (CRB)-based screening criterion under the adopted sensing model. The retained candidates are then ranked according to their recovered transmit power.
    
        \item For each retained grouping pattern, we formulate transmit-power minimization subject to CU signal-to-interference-plus-noise ratio (SINR), target signal-to-clutter-plus-noise ratio (SCNR), illumination, and response-correlation constraints, and solve the resulting fixed-partition problem using SDR-based alternating optimization (AO) and successive convex approximation (SCA). We distinguish the conservative approximation of the sensing constraints, semidefinite rank relaxation, and subsequent rank-one recovery, and verify the recovered solution against the original physical quality-of-service constraints.
    \end{itemize}
    
    \emph{Organization:} Section II presents the system model and formulates the joint design problem. Section III develops the S-CSI-guided task-dependent grouping strategy. Section IV presents the fast-timescale active--passive beamforming algorithm. Section V provides numerical results, and Section VI concludes the paper. 
    
    \emph{Notation:} $(\cdot)^T$, $(\cdot)^H$, $\operatorname{Tr}(\cdot)$, $\operatorname{rank}(\cdot)$, $\operatorname{diag}(\cdot)$, and $\operatorname{vec}(\cdot)$ denote transpose, Hermitian transpose, trace, rank, diagonalization, and vectorization, respectively. The operators $\odot$, $\otimes$, and $\mathbb E\{\cdot\}$ denote the Hadamard product, Kronecker product, and expectation, respectively. Identity and zero matrices of compatible dimensions are denoted by $\mathbf I$ and $\mathbf 0$.
    
    \section{System Model and Problem Formulation}
    \subsection{System Model}
    \begin{figure}[htbp]
        \centerline{\includegraphics[width=\columnwidth]{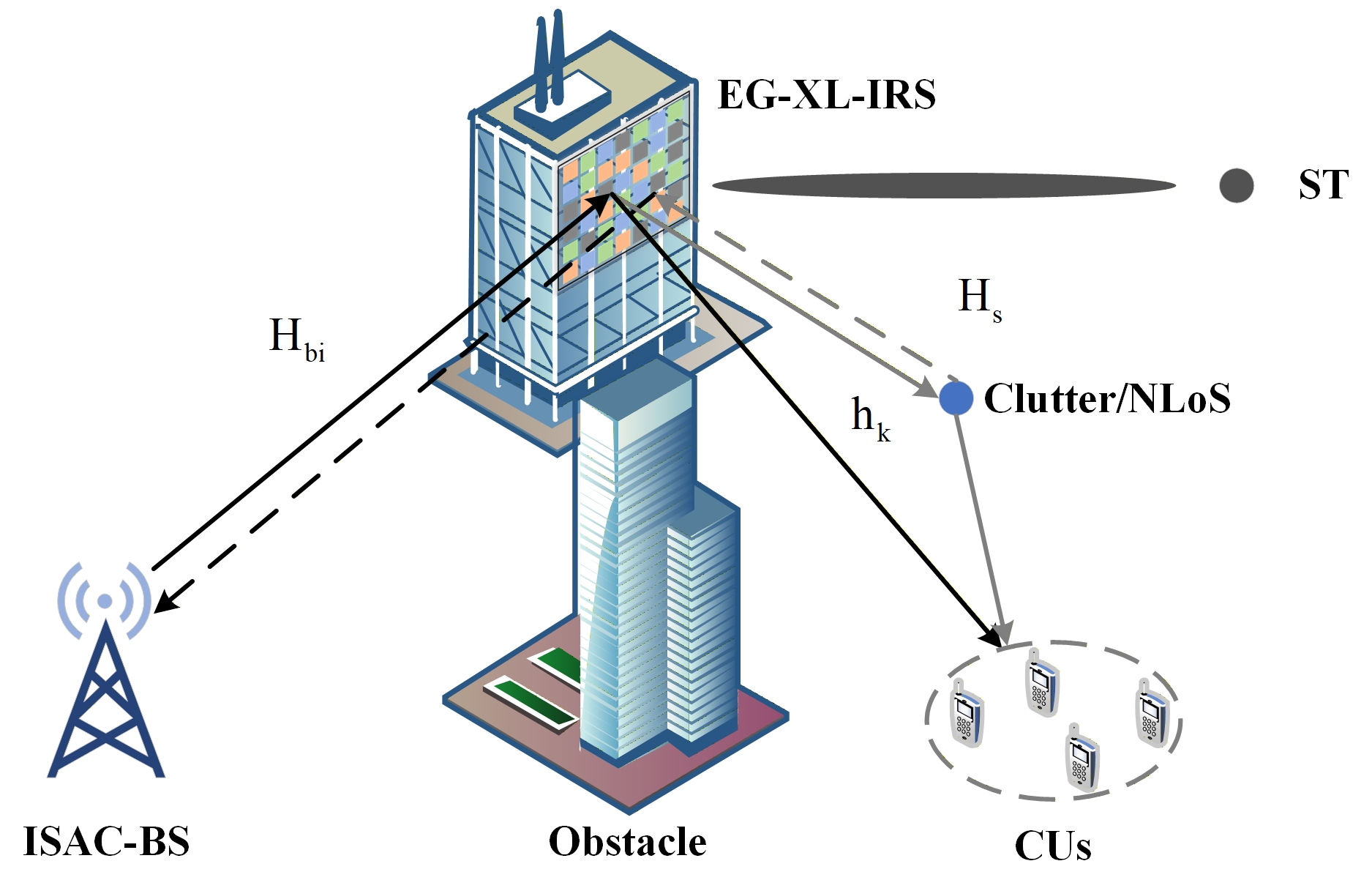}}
            \caption{EG-XL-IRS-assisted multiuser ISAC system with one desired target and non-target clutter scatterers.}
        \label{fig:1}
    \end{figure}
    As shown in Fig.~\ref{fig:1}, an ISAC-BS equipped with an $M$-antenna uniform linear array (ULA) serves $K$ single-antenna CUs while sensing one desired target comprising $T\geq1$ scattering centers. The environment also contains $L$ non-target scatterers whose echoes are treated as clutter. The CU, desired-scatterer, and clutter index sets are denoted by $\mathcal K$, $\mathcal T$, and $\mathcal L$, respectively. Thus, $T=1$ corresponds to a point target, whereas $T>1$ represents an extended target. For clarity, the direct BS--CU and BS--target links are assumed blocked, and an EG-XL-IRS with $N$ reflecting elements provides the controllable reflected paths. Let $\bm G\in\{0,1\}^{Q\times N}$ denote the grouping matrix, where $G_{q,n}=1$ indicates that element $n$ belongs to group $q$. All elements in group $q$ share the reflection coefficient $v_q=e^{j\theta_q}$, and $\bm v=[v_1,\ldots,v_Q]^T\in\mathbb C^Q$. The resulting element-wise reflection matrix is
    \begin{equation}
        \label{eq:1}
        \bm{\Phi} = \operatorname{diag}\left\{ \bm{G}^T\bm{v} \right\},
    \end{equation}

    Following \cite{10440056}, the ISAC-BS transmits both communication symbols and a dedicated sensing waveform so that the transmit DoFs can also be used for sensing. The transmitted signal is
    \begin{equation}
        \label{eq:2}
        \bm{x} = \bm{W}_s\mathbf{s} + \bm{W}_c\mathbf{c},
    \end{equation}
    where $\bm W_s\in\mathbb C^{M\times M}$ is the sensing beamforming matrix and $\bm W_c=[\bm w_1,\ldots,\bm w_K]\in\mathbb C^{M\times K}$ is the communication beamforming matrix. The sensing vector $\mathbf s\in\mathbb C^M$ satisfies $\mathbb E\{\mathbf s\}=\mathbf0$ and $\mathbb E\{\mathbf s\mathbf s^H\}=\mathbf I_M$, while $\mathbf c=[c_1,\ldots,c_K]^T\sim\mathcal{CN}(\mathbf0,\mathbf I_K)$ collects the communication symbols. We assume $\mathbf s$ and $\mathbf c$ are statistically independent, i.e., $\mathbb E\{\mathbf s\mathbf c^H\}=\mathbf0$.
    
    The resulting transmit covariance matrix is
    \begin{equation}
        \label{eq:3}
        \bm{R}_x \triangleq \mathbb E\left[ {\bm{x}{\bm{x}^H}} \right] = \bm{R}_s + \sum_{k \in \mathcal{K}} \bm{R}_k,
    \end{equation}
    where $\bm{R}_s = \bm{W}_s\bm{W}_s^H$ and $\bm{R}_k = \bm{w}_k \bm{w}^H_k$.

    Assuming half-wavelength spacing at both the ISAC-BS and the EG-XL-IRS, the response of an $N$-element ULA is \cite{10078317}
    \begin{equation}
        \label{eq:4}
        {\mathbf{a}_N}\left( \theta \right) = \left[ 1, e^{ - j\pi\sin{\theta}}, \ldots ,e^{ - j\pi(N-1)\sin{\theta}} \right]^T,
    \end{equation}
    where $\theta\in(-\pi/2,\pi/2)$ denotes the azimuth angle relative to array broadside. We adopt a narrowband far-field plane-wave model and assume that the angular supports of the target and clutter are available from prior acquisition or tracking. Near-field propagation, mutual coupling, and blind initial acquisition are outside the present scope. \footnote{The far-field model is a deliberate abstraction used to isolate the grouping-induced gain--rank and dimensionality-reduction effects. Thus, $N=4096$ is used to stress the element-domain acquisition/control dimension, not to assert that a half-wavelength 4096-element physical aperture at the stated geometry necessarily satisfies the Fraunhofer far-field condition. }

    As illustrated in Fig.~\ref{fig:1}, let $\mathbf H_{bi}\in\mathbb C^{M\times N}$, $\mathbf h_k\in\mathbb C^N$, and $\mathbf H_s\in\mathbb C^{N\times N}$ denote the BS--IRS channel, IRS--CU-$k$ channel, and IRS-domain sensing response, respectively. Based on \eqref{eq:4}, we model these quantities as \cite{10762897}
    \begin{equation}
        \label{eq:5}
        \begin{array}{*{20}{l}}
            &\mathbf{H}_{bi} = \underbrace{\alpha\sqrt {\frac{\kappa_{bi}}{{1 + \kappa_{bi}}}} \bar{\mathbf{H}}_{bi}}_{\mathrm{LoS}} + \underbrace{\alpha\sqrt {\frac{1}{{1 + \kappa_{bi}}}} \tilde{\mathbf{H}}_{bi}}_{\mathrm{NLoS}}, \\
            &\mathbf{h}_{k} = \underbrace{\beta_k\sqrt {\frac{\kappa_{k}}{{1 + \kappa_{k}}}} \bar{\mathbf{h}}_{k}}_{\mathrm{LoS}} + \underbrace{\beta_k\sqrt {\frac{1}{{1 + \kappa_{k}}}} \tilde{\mathbf{h}}_{k}}_{\mathrm{NLoS}},\quad k \in \mathcal{K},\\
            &\mathbf{H}_{s} = \underbrace{\sum_{t \in \mathcal{T}}\gamma_{t}{\mathbf{a} _{N}}\left( \theta_{s}^t \right){\mathbf{a} _{N}^H}\left( \theta_{s}^t \right)}_{\mathrm{Target}} + \underbrace{\sum_{l \in \mathcal{L}} {\gamma_l {\mathbf{a} _{N}}\left( \theta_{s}^l \right){\mathbf{a} _{N}^H}\left( \theta_{s}^l \right)}}_{\mathrm{Clutter}},
        \end{array}
    \end{equation}
    where $\bar{\mathbf{H}}_{bi} = {\mathbf{a} _{M}}\left( \theta_{bi} \right){\mathbf{a} _N^H}{\left( \phi_{bi} \right)}$, $\bar{\mathbf{h}}_{k} = {\mathbf{a} _N}{{\left( \phi_{k} \right)}}$, $\operatorname{vec}(\tilde{\mathbf{H}}_{bi}) \sim \mathcal{CN}(\mathbf{0}, \mathbf{I}_{MN})$ and $\tilde{\mathbf{h}}_{k} \sim \mathcal{CN}(\mathbf{0}, \mathbf{I}_N)$. $\alpha$, $\theta_{bi}$, and $\phi_{bi}$ denote the complex large-scale gain, angle of departure, and angle of arrival of the BS--IRS LoS path, respectively. Similarly, $\beta_k$ and $\phi_k$ specify the large-scale gain and LoS direction of CU $k$. For desired scatterer $t$, $|\gamma_t|$ denotes its effective reflection magnitude (an radar-cross-section (RCS)-related coefficient under the adopted normalization) and $\theta_s^t$ its direction; for clutter scatterer $l$, $\gamma_l\sim\mathcal{CN}(0,\delta_l^2)$ models a random scattering coefficient at angular location $\theta_s^l$. The dominant angles, path losses, Rician factors, desired-scatterer strengths $\{|\gamma_t|\}$, and clutter variances $\{\delta_l^2\}$ are treated as slowly varying S-CSI. Instantaneous NLoS coefficients and scattering phases/realizations are handled on the fast timescale and are not assumed known when constructing $\bm G$.

    Because a passive IRS cannot separately estimate its incident and reflected links, the ISAC-BS operates on cascaded CSI. We assume that the deliberately deployed BS--IRS and IRS--CU/target links contain exploitable LoS components. Separating the scattering magnitude from its phase, the one-way factors used to construct the cascaded CU, desired-scatterer, and clutter responses are
    \begin{equation}
        \label{eq:7}
        \begin{split}
            \mathbf{C}_{k} =& \mathbf{H}_{bi} \operatorname{diag}\left\{\mathbf{h}_k\right\}, \\ 
            \mathbf{C}_{s,t} =& \underbrace{\sqrt{\left | \gamma_{t} \right |}\mathbf{H}_{bi} \operatorname{diag}\left\{\mathbf{a} _N\left( {\theta^{t}_{s}} \right)\right\}}_{\mathrm{Cascaded\ Target\ Component}}, \\ 
            \mathbf{C}_{s,l} =& \underbrace{\sqrt{\left | \gamma_{l} \right |}\mathbf{H}_{bi} \operatorname{diag}\left\{\mathbf{a} _N\left( {\theta^{l}_{s}} \right)\right\}}_{\mathrm{Cascaded\ Clutter\ Component}}, \\
            &k \in \mathcal{K}, t \in \mathcal{T}, l \in \mathcal{L},
        \end{split}
    \end{equation}
    where $\mathbf C_k$ is the cascaded BS--IRS--CU-$k$ channel. The matrices $\mathbf C_{s,t}$ and $\mathbf C_{s,l}$ are one-way factors whose outer products generate, respectively, the desired-target and clutter round-trip responses after multiplication by the corresponding scattering phase. We refer to $\mathbf C_{s,l}$ as a \emph{cascaded clutter component}: this terminology identifies its role in the end-to-end sensing channel and does not imply that the clutter is spatially isotropic or coherent with the desired target.

    Grouping reduces the dimensions of cascaded-CSI acquisition and online phase optimization while retaining deterministic structures relevant to communication and desired-target sensing. The partition uses only slow angular/statistical information, not realization-specific NLoS samples or clutter phases; angular support is assumed available from prior acquisition or tracking.
        
    \subsection{Problem Formulation}
    For the considered EG-XL-IRS-assisted system, the post-grouping cascaded channels associated with $\mathbf C_k$, $\mathbf C_{s,t}$, and $\mathbf C_{s,l}$ are $\mathbf C_{g,k}\triangleq\mathbf C_k\bm G^T\in\mathbb C^{M\times Q}$, $\mathbf C_{gs,t}\triangleq\mathbf C_{s,t}\bm G^T\in\mathbb C^{M\times Q}$, and $\mathbf C_{gs,l}\triangleq\mathbf C_{s,l}\bm G^T\in\mathbb C^{M\times Q}$, respectively. Thus, the dimension of each cascaded channel seen by the online optimizer is reduced from $N$ to $Q$.

    The protocol operates on two timescales. Slowly varying path losses, Rician factors, dominant angles, and scattering-power statistics determine $\bm G$, whereas the ISAC-BS estimates the $Q$-dimensional post-grouping responses and updates the group phases within each coherence interval. Under the adopted group-wise estimation protocol, the IRS-related pilot/configuration count scales with $Q$ rather than $N$, yielding an overhead--control-resolution tradeoff.

    For the communication link, the received signal at CU $k$ is given by
    \begin{equation}
        \label{eq:8}
        y_k = \mathbf{h}_{g,k}^H\bm{x} + n_k,
    \end{equation}
    where $\mathbf h_{g,k}=\mathbf C_{g,k}\bm v$ is the effective channel of CU $k$ and $n_k\sim\mathcal{CN}(0,\sigma_k^2)$ is receiver noise.

    From \eqref{eq:3} and \eqref{eq:8}, the received SINR at CU $k$ is given by
    \begin{equation}
        \label{eq:9}
        \mathrm{SINR}_{k} = \frac{\operatorname{Tr}\left(\bm R_k \mathbf H_{g,k}\right)}{\sum_{i\in\mathcal K\setminus\{k\}}\operatorname{Tr}\left(\bm R_i \mathbf H_{g,k}\right) + \operatorname{Tr}\left(\bm R_s \mathbf H_{g,k}\right) + \sigma_k^2},
    \end{equation}
    where $\mathbf{H}_{g,k} = \mathbf{h}_{g,k}\mathbf{h}_{g,k}^H$. 

    The dedicated sensing waveform contributes probing energy but appears as interference at the CUs; we assume that the CUs do not cancel this waveform, so its covariance must be optimized jointly with the communication covariances.
    
    For the sensing link, the received echo at the ISAC-BS is modeled as
    \begin{equation}
        \label{eq:10}
        \mathbf y_s
        =\underbrace{\mathbf H_{gt}\bm x}_{\text{desired-target echo}}
        +\underbrace{\mathbf H_{gc}\bm x}_{\text{clutter echo}}
        +\mathbf n_s,
    \end{equation}
    where $\mathbf n_s\sim\mathcal{CN}(\mathbf0,\sigma_s^2\mathbf I_M)$ is the receiver noise. Define $\mathbf h_{gt,t}=\mathbf C_{gs,t}\bm v$ and $\mathbf h_{gc,l}=\mathbf C_{gs,l}\bm v$. The per-scatterer round-trip matrices and the corresponding aggregate desired-target and clutter responses are
    \begin{equation}
        \begin{aligned}
        \mathbf H_{gt,t}&=\gamma_{s,t}\mathbf h_{gt,t}\mathbf h_{gt,t}^H,
        &\mathbf H_{gc,l}&=\gamma_{s,l}\mathbf h_{gc,l}\mathbf h_{gc,l}^H,\\
        \mathbf H_{gt}&=\sum_{t\in\mathcal T}\mathbf H_{gt,t},
        &\mathbf H_{gc}&=\sum_{l\in\mathcal L}\mathbf H_{gc,l}.
        \end{aligned}
        \label{eq:round_trip_components}
    \end{equation}
    where $\gamma_{s,t}=\gamma_t/|\gamma_t|$ and $\gamma_{s,l}=\gamma_l/|\gamma_l|$ retain the scattering phases, with nonzero scattering coefficients understood.

    Using the clutter-whitened receive model in \cite{9913311}, the target SCNR is
    \begin{equation}
        \label{eq:11}
        \mathrm{SCNR}=\operatorname{Tr}\!\left(
        \mathbf H_{gt}\bm R_x\mathbf H_{gt}^H\mathbf D^{-1}\right),
    \end{equation}
    where $\mathbf D=\mathbf H_{gc}\bm R_x\mathbf H_{gc}^H+\sigma_s^2\mathbf I_M\succ\mathbf0$ is the conditional clutter-plus-noise covariance matrix for the estimated fast-timescale clutter response.

    Aggregate SCNR alone does not guarantee sufficient excitation of every desired scattering center. We use the received energy associated with scattering center $t$, defined as
    \begin{equation}
        \mathcal E_t(\mathbf R_x,\bm v)
        =\operatorname{Tr}\!\left(
        \mathbf H_{gt,t}\mathbf R_x\mathbf H_{gt,t}^H\right).
        \label{eq:v22_scatterer_energy}
    \end{equation}
    The constraints $\mathcal E_t\geq\underline e_t$ prevent the optimizer from satisfying the aggregate SCNR primarily through the strongest scattering center.

    To limit mutual correlation among desired-scatterer responses, define the cross-correlation budget (CCB) as
    \begin{equation}
        \operatorname{CCB}
        =\sum_{\substack{t,u\in\mathcal T, \ t<u}}
        \left|
        \operatorname{Tr}\!\left(
        \mathbf H_{gt,t}\mathbf R_x\mathbf H_{gt,u}^H\right)
        \right|.
        \label{eq:v22_ccb}
    \end{equation}
    For $T=1$, the sum in \eqref{eq:v22_ccb} is empty and the CCB constraint is inactive. Because the unnormalized CCB has channel- and power-dependent units, $\epsilon_s$ is normalized using the same reference channel gain and noise power adopted in the simulations.
    
    We first state the full joint design problem, in which the ISAC-BS transmit power is minimized over the active beamformers, group-wise phases, and binary grouping pattern:
    \begin{subequations}
        \label{eq:13}
        \begin{align}
            {{P_0}:} \min_{\bm W_s,\bm W_c,\bm G,\bm v} &\operatorname{Tr}(\bm R_x), \nonumber\\
            \mathrm{s.t.} \quad &\mathrm{SINR}_k \geq \Gamma_k,\quad k \in \mathcal K, \label{eq:13A}\\
            &\mathrm{SCNR} \geq \Gamma_s, \label{eq:13B}\\
            &\mathcal E_t \geq \underline e_t,\quad t \in \mathcal T, \label{eq:13C}\\
            &\operatorname{CCB} \leq \epsilon_s, \label{eq:13D}\\
            &|v_q| = 1,\quad q=1,\ldots,Q, \label{eq:13E}\\
            &G_{q,n}\in\{0,1\},\quad \sum_{q=1}^QG_{q,n}=1, \nonumber\\
            &\sum_{n=1}^NG_{q,n}\geq1,\quad q=1,\ldots,Q. \label{eq:13F}
        \end{align}
    \end{subequations}
    Here, $\Gamma_k$ and $\Gamma_s$ denote the required SINR of CU $k$ and the required target SCNR, respectively. Constraints \eqref{eq:13C} and \eqref{eq:13D} protect the desired-scatterer illumination and response correlation, \eqref{eq:13E} enforces unit-modulus group coefficients, and \eqref{eq:13F} assigns every reflecting element to exactly one nonempty group.

    Problem $P_0$ is a mixed-integer nonconvex program due to the binary grouping variables, unit-modulus constraints, coupling between active and passive variables, and quartic dependence of the round-trip sensing terms on $\bm v$. Direct global optimization is therefore impractical in the considered XL-IRS regime. We instead adopt a two-timescale candidate-restricted approach. Section III uses S-CSI to construct and screen a finite set of grouping candidates. For each retained candidate, the $Q$-dimensional post-grouping cascaded channels are estimated, after which Section IV alternately optimizes the transmit covariances and group-wise phases. The final grouping is selected according to the recovered transmit power among candidates satisfying the prescribed feasibility and rank/diversity criteria.
    
    The slow-timescale design relies on path losses, Rician factors, dominant angles, and scattering-power statistics, avoiding realization-wise optimization over all $N$ IRS elements. The fast-timescale design still requires instantaneous post-grouping cascaded CSI; grouping reduces the channel-estimation and optimization dimensions but does not remove the need for CSI acquisition. Hence, the proposed framework is a task-aware dimensionality-reduction strategy rather than a CSI-free passive-beamforming scheme.
    
    \section{Task-Adaptive S-CSI-Guided Grouping for Point and Extended Targets}
    \label{sec:task_adaptive_grouping}
    
    We construct grouping candidates from slowly varying S-CSI. For $T=1$, only communication-channel separability must be protected against excessive common-mode concentration. For $T>1$, the grouping must additionally prevent the desired-scatterer responses from collapsing into a nearly one-dimensional subspace; hence the extended-target branch includes per-scatterer illumination and response-diversity safeguards.
        
    \subsection{Common Coherent-Gain Benefit and Common-Mode Rank Cost}
    \label{subsec:common_effects}
    
    The post-grouping cascaded communication channel of CU $k$ is decomposed as
    \begin{equation}
      \mathbf C_{g,k}
      =\mathbf C_k\bm G^T
      =\overline{\mathbf C}_{g,k}+\widetilde{\mathbf C}_{g,k},
      \qquad k\in\mathcal K,
      \label{eq:Cgk_task_decomposition}
    \end{equation}
    where
    \begin{equation}
      \overline{\mathbf C}_{g,k}
      =\eta_k\mathbf a_M(\theta_{bi})\mathbf b_k^H\bm G^T,
      \quad
      \mathbf b_k=\mathbf a_N(\phi_{bi})\odot\mathbf a_N^*(\phi_k).
      \label{eq:Cgk_task_los}
    \end{equation}
    Likewise, for desired scattering center $t\in\mathcal T$ and clutter scatterer $l\in\mathcal L$, we write
    \begin{align}
      \mathbf C_{gs,t}
      &=\overline{\mathbf C}_{gs,t}+\widetilde{\mathbf C}_{gs,t},
      &
      \overline{\mathbf C}_{gs,t}
      &=\eta_t\mathbf a_M(\theta_{bi})\mathbf b_t^H\bm G^T,
      \label{eq:Cgst_task_decomposition}\\
      \mathbf C_{gs,l}
      &=\mathbf C_{s,l}\bm G^T,
      &
      \mathbf b_t
      &=\mathbf a_N(\phi_{bi})\odot\mathbf a_N^*(\theta_s^t).
      \label{eq:Cgsl_task_decomposition}
    \end{align}
    The large-scale gains and Rician factors are retained in $\eta_k=\alpha\beta_k\sqrt{\frac{\kappa_{bi}\kappa_k}{(1+\kappa_{bi})(1+\kappa_k)}}$ and $\eta_t=\alpha\sqrt{\frac{\kappa_{bi}}{1+\kappa_{bi}}|\gamma_t|}$. Retaining these factors is essential because the slow-timescale grouping should distinguish a distant CU or weak desired scattering center from a nearby strong component. We do not assign a coherent-gain utility to clutter in the outer grouping objective: $\mathbf C_{gs,l}$ may remain directionally structured, but it represents an undesired and realization-dependent component whose impact is handled by the fast-timescale SCNR constraint and the final feasibility audit.
    
    Let $\mathcal S_q$ be the element set of group $q$ and $\mu_q=|\mathcal S_q|$. Since the groups are disjoint, we have
    \begin{equation}
      \bm G\bm G^T
      =\operatorname{diag}(\mu_1,\ldots,\mu_Q),
      \qquad \sum_{q=1}^Q\mu_q=N.
      \label{eq:GGT_task}
    \end{equation}
    For an isotropic diffuse component $\widetilde{\mathbf c}\sim\mathcal{CN}(\mathbf0,\sigma_c^2\mathbf I_N)$, it follows that
    \begin{equation}
      \mathbb E\!\left\{\|\bm G\widetilde{\mathbf c}\|^2\right\}
      =\sigma_c^2\sum_{q=1}^Q\mu_q=N\sigma_c^2.
      \label{eq:diffuse_power_task}
    \end{equation}
        Thus, grouping preserves the total average power of the modeled i.i.d. isotropic NLoS component; this identity does not apply to a directional clutter scatterer in \eqref{eq:5}. If the deterministic phases of component $i$ are approximately aligned within each group, then
        \begin{equation}
      \|\bm G\overline{\mathbf c}_i\|^2
      \approx |\eta_i|^2\sum_{q=1}^Q\mu_q^2.
      \label{eq:coherent_gain_task}
    \end{equation}
    where $\overline{\mathbf c}_i=\eta_i\mathbf b_i$. For equal-size groups, the coherent deterministic term in \eqref{eq:coherent_gain_task} scales as $N^2/Q$, whereas the diffuse-power term in \eqref{eq:diffuse_power_task} scales as $N$. Phase-consistent grouping can therefore increase the deterministic-to-diffuse power ratio of selected components without reducing the absolute average diffuse power.
        
    Let $\overline{\mathbf v}=\bm G^T\mathbf v$ and collect the effective CU and desired-scatterer one-way channels as
    \begin{equation}
      \mathbf H_c=[\mathbf h_{g,1},\ldots,\mathbf h_{g,K}],
      \quad
      \mathbf H_{\mathcal T}=[\mathbf h_{gt,1},\ldots,\mathbf h_{gt,T}].
      \label{eq:Hc_HT_task}
    \end{equation}
    Their deterministic LoS components satisfy
    \begin{align}
      \overline{\mathbf H}_c
      &=\mathbf a_M(\theta_{bi})[q_1,\ldots,q_K],
      &q_k&=\eta_k\mathbf b_k^H\overline{\mathbf v},
      \label{eq:Hc_los_task}\\
      \overline{\mathbf H}_{\mathcal T}
      &=\mathbf a_M(\theta_{bi})[s_1,\ldots,s_T],
      &s_t&=\eta_t\mathbf b_t^H\overline{\mathbf v}.
      \label{eq:HT_los_task}
    \end{align}
    Consequently,
    \begin{equation}
      \operatorname{rank}(\overline{\mathbf H}_c)\le 1,
      \qquad
      \operatorname{rank}(\overline{\mathbf H}_{\mathcal T})\le 1.
      \label{eq:common_los_rank_task}
    \end{equation}
    For a nonzero matrix $\mathbf X$ with singular values $\{\sigma_j(\mathbf X)\}$, define
    \begin{align}
      p_j(\mathbf X)
      &=\frac{\sigma_j^2(\mathbf X)}{\|\mathbf X\|_F^2},
      \label{eq:singular_energy_task}\\
      r_{\rm eff}(\mathbf X)
      &=\exp\!\left[-\sum_jp_j(\mathbf X)\log p_j(\mathbf X)\right],
      \label{eq:effective_rank_task}
    \end{align}
    where the convention $0\log 0=0$ is used.
    
    \emph{Proposition 1 (effective-rank collapse under common-mode domination):}
    Let $\mathbf H_i=\overline{\mathbf H}_i+\widetilde{\mathbf H}_i$ for $i\in\{c,\mathcal T\}$, where $\operatorname{rank}(\overline{\mathbf H}_i)\le1$ and $\sigma_1(\overline{\mathbf H}_i)>0$. For fixed matrix dimensions, if
    \begin{equation}
      \epsilon_i\triangleq
      \frac{\|\widetilde{\mathbf H}_i\|_2}
           {\sigma_1(\overline{\mathbf H}_i)}\longrightarrow0,
      \label{eq:epsilon_task}
    \end{equation}
    then
    \begin{equation}
      \frac{\sigma_2(\mathbf H_i)}{\sigma_1(\mathbf H_i)}\longrightarrow0,
      \qquad r_{\rm eff}(\mathbf H_i)\longrightarrow1.
      \label{eq:rank_collapse_task}
    \end{equation}

    \begin{IEEEproof}
        By the Eckart--Young theorem, $\sigma_2(\mathbf H_i)\le\|\widetilde{\mathbf H}_i\|_2$, while the reverse triangle inequality gives $\sigma_1(\mathbf H_i)\ge\sigma_1(\overline{\mathbf H}_i)-\|\widetilde{\mathbf H}_i\|_2$. Hence, for $\epsilon_i<1$, $\sigma_2(\mathbf H_i)/\sigma_1(\mathbf H_i)\le\epsilon_i/(1-\epsilon_i)\to0$. Because the matrix dimensions are fixed, all singular values beyond the first are likewise bounded by the residual spectral norm; their normalized energy therefore vanishes, yielding $r_{\rm eff}(\mathbf H_i)\to1$.
	\end{IEEEproof}
    
    Proposition 1 has different implications for the two sensing tasks. For $T=1$, $\mathbf H_{\mathcal T}$ contains a single column and is rank one by construction, so coherent target enhancement cannot eliminate an additional desired-target dimension. The communication matrix $\mathbf H_c$, however, should retain multiple energetic singular modes whenever the CUs rely on spatial separation. For $T>1$, $r_{\rm eff}(\mathbf H_{\mathcal T})\approx1$ indicates that the desired-scatterer energy is concentrated in one dominant response mode despite the presence of multiple physical scattering centers. In this regime, aggregate echo gain and useful target-response diversity can diverge. We therefore use effective rank together with the dominant-mode energy fraction, rather than algebraic rank alone, when screening grouping candidates.
        
    \subsection{Point-Target Branch: Weak-Link Gain Enhancement With Communication-Rank Protection}
    \label{subsec:point_target_branch}
    
    For $T=1$, denote the point target by $p$. Let $\mathcal K_{\rm w}\subseteq\mathcal K$ contain the path-loss-limited CUs identified from their cascaded large-scale gains; it contains at least the weakest CU and may include several distant CUs. Define
    \begin{equation}
      \mathbf U_k=\overline{\mathbf C}_k^H\overline{\mathbf C}_k,
      \qquad
      \mathbf U_p=\overline{\mathbf C}_{s,p}^H\overline{\mathbf C}_{s,p},
      \label{eq:U_point}
    \end{equation}
    where $\overline{\mathbf C}_k=\eta_k\mathbf a_M(\theta_{bi})\mathbf b_k^H$ and $\overline{\mathbf C}_{s,p}=\eta_p\mathbf a_M(\theta_{bi})\mathbf b_p^H$. For $0<r<1$, the normalized deterministic gains are
    \begin{align}
      g_k^{\rm pt}(\overline{\mathbf v};r)
      &=\frac{\overline{\mathbf v}^H\mathbf U_k\overline{\mathbf v}}
      {r\Gamma_k\sigma_k^2},
      &&k\in\mathcal K,
      \label{eq:gk_point}\\
      g_p^{\rm pt}(\overline{\mathbf v};r)
      &=\frac{\overline{\mathbf v}^H\mathbf U_p\overline{\mathbf v}}
      {(1-r)\Gamma_s\sigma_s^2}.
      \label{eq:gp_point}
    \end{align}
    Because $|\eta_k|^2$ is retained in $\mathbf U_k$, a path-loss-limited CU typically attains a smaller normalized deterministic gain and is therefore naturally emphasized by the max--min objective.
    
    To avoid rewarding an already strong common LoS mode without limit, we define the bounded utility
    \begin{equation}
      \psi_i(g)=1-\exp(-g/\chi_i),
      \qquad \chi_i>0.
      \label{eq:bounded_utility_task}
    \end{equation}
    The point-target candidate is generated by
    \begin{align}
      \mathcal P_{\rm PT}(r):\quad
      \max_{\overline{\mathbf v}}\quad
      &\min_{i\in\mathcal I_{\rm PT}}
      \psi_i\!\left(g_i^{\rm pt}(\overline{\mathbf v};r)\right)
      \label{prob:point_gain}\\
      \text{s.t.}\quad
      &|\overline v_n|=1,\quad n=1,\ldots,N,\nonumber\\
      &\psi_k(g_k^{\rm pt})\ge \underline u_k,
      \quad k\in\mathcal K\setminus\mathcal K_{\rm w},
      \nonumber
    \end{align}
    where $\mathcal I_{\rm PT}=\mathcal K_{\rm w}\cup\{p\}$. The floors $\{\underline u_k\}$ prevent the weak-link objective from severely degrading the remaining CUs. In an approximately noise-limited regime, a useful power scaling is
    \begin{equation}
      p_k^{\rm lb}\approx
      \frac{\Gamma_k\sigma_k^2}{\|\mathbf h_{g,k}\|^2}.
      \label{eq:point_far_user_power}
    \end{equation}
    Equation~\eqref{eq:point_far_user_power} motivates using the weak-CU deterministic gain as an outer-layer proxy in the noise-limited regime, while the final performance is governed by the coupled SINR/SCNR design. Communication-rank protection is enforced explicitly during candidate screening rather than inferred from gain alone.
        
    \subsection{Extended-Target Branch: Per-Scatterer Illumination and Angular-Separability Preservation}
    \label{subsec:extended_target_branch}
    
    For $T>1$, the target cannot be replaced by a single representative direction. Define $\mathbf U_{s,t}=\overline{\mathbf C}_{s,t}^H\overline{\mathbf C}_{s,t}$ and
    \begin{equation}
      g_t^{\rm et}(\overline{\mathbf v};r)
      =\frac{\overline{\mathbf v}^H\mathbf U_{s,t}\overline{\mathbf v}}
      {(1-r)\nu_t\Gamma_s\sigma_s^2},
      \quad t\in\mathcal T,
      \label{eq:gt_extended}
    \end{equation}
    where $\nu_t>0$ and $\sum_{t\in\mathcal T}\nu_t=1$ specify the desired illumination profile. The user gains $g_k^{\rm et}$ are defined as in \eqref{eq:gk_point}. A gain-balanced element-wise candidate is first obtained from
    \begin{align}
      \max_{|\overline v_n|=1}\quad
      &\min_{i\in\mathcal I_{\rm ET}}
      \psi_i\!\left(g_i^{\rm et}(\overline{\mathbf v};r)\right)
      \label{prob:extended_gain_seed}\\
      \text{s.t.}\quad
      &\psi_t(g_t^{\rm et})\ge\underline u_t,
      \quad t\in\mathcal T,\nonumber\\
      &\psi_k(g_k^{\rm et})\ge\underline u_k,
      \quad k\in\mathcal K\setminus\mathcal K_{\rm w},\nonumber
    \end{align}
    where $\mathcal I_{\rm ET}=\mathcal K_{\rm w}\cup\mathcal T$. The constraints for all $t$ prevent the outer gain objective from concentrating only on the strongest scattering center. They are S-CSI-based surrogates used for candidate generation; the exact received-energy constraints are enforced again in the fast-timescale design.
    
    Gain balancing alone does not protect angular separability. For a recovered grouping matrix $\bm G$, we define the group-domain signature of scattering center $t$ as
    \begin{equation}
      \mathbf z_t(\bm G)
      =\operatorname{vec}(\overline{\mathbf C}_{s,t}\bm G^T)
      =\eta_t(\bm G\mathbf b_t^*)\otimes\mathbf a_M(\theta_{bi}),
      \label{eq:group_domain_signature}
    \end{equation}
    and its normalized version $\widehat{\mathbf z}_t=\mathbf z_t/\|\mathbf z_t\|$. Let $\widehat{\mathbf Z}_{\mathcal T}=[\widehat{\mathbf z}_1,\ldots,\widehat{\mathbf z}_T]$. We use the worst-case signature coherence
    \begin{equation}
      \mu_{\rm ang}(\bm G)
      =\max_{t\ne u}
      |\widehat{\mathbf z}_t^H\widehat{\mathbf z}_u|^2,
      \label{eq:angular_coherence}
    \end{equation}
    and the normalized log-determinant diversity
    \begin{equation}
      D_{\rm ang}(\bm G)
      =\frac{1}{T}\log\det\!\left(
      \varepsilon\mathbf I_T+\widehat{\mathbf Z}_{\mathcal T}^H
      \widehat{\mathbf Z}_{\mathcal T}\right),
      \quad \varepsilon>0.
      \label{eq:angular_logdet}
    \end{equation}
    The coherence term penalizes the most ambiguous pair, whereas the log-determinant rewards global signature diversity. Both are S-CSI-based structural surrogates; neither alone guarantees a particular angular-estimation accuracy.
        
    After the gain seed is quantized into balanced groups, the extended-target grouping is refined through pairwise element swaps that monotonically increase
    \begin{equation}
      J_{\rm ET}(\bm G,\mathbf v)
      =F_{\tau}^{\rm ET}(\bm G^T\mathbf v)
      +\lambda_DD_{\rm ang}(\bm G)
      -\lambda_\mu\mu_{\rm ang}(\bm G),
      \label{eq:extended_refinement_objective}
    \end{equation}
    subject to the gain floors in \eqref{prob:extended_gain_seed}. Here, $F_{\tau}^{\rm ET}$ is the smooth approximation of the minimum bounded utility, while $\lambda_D,\lambda_\mu\ge0$ weight the diversity terms. The group centroids are updated after every accepted swap. Because each swap preserves group sizes and is accepted only when it strictly increases \eqref{eq:extended_refinement_objective}, the finite assignment process terminates at a swap-wise local optimum.
    
    For estimator-aware screening, let $\mathbf H_{gt}(\boldsymbol\theta)$ denote the aggregate desired-target round-trip channel, $\dot{\mathbf H}_t=\partial\mathbf H_{gt}/\partial\theta_s^t$, $L_p$ the number of independent probing snapshots, and $\mathbf D$ the clutter-plus-noise covariance. Under the adopted Gaussian receive model, with $\mathbf D$ treated as independent of the target-angle parameters, a conditional angle Fisher information matrix (FIM) for a candidate with optimized $\mathbf R_x$ and $\bm v$ is
    \begin{equation}
      [\mathbf J_{\boldsymbol\theta}]_{t,u}
      =2L_p\operatorname{Re}\!\left\{
      \operatorname{Tr}\!\left(
      \mathbf D^{-1}\dot{\mathbf H}_t\mathbf R_x\dot{\mathbf H}_u^H
      \right)\right\}.
      \label{eq:angle_fim}
    \end{equation}
    If the complex target-scattering coefficients collected in $\boldsymbol\gamma_{\mathcal T}$ are unknown nuisance parameters and the corresponding real-parameter block $\mathbf J_{\gamma\gamma}$ is nonsingular, the angle-only equivalent FIM is obtained via the Schur complement
    \begin{equation}
      \mathbf J_{\boldsymbol\theta,\rm nuis}
      =\mathbf J_{\theta\theta}
      -\mathbf J_{\theta\gamma}\mathbf J_{\gamma\gamma}^{-1}
      \mathbf J_{\gamma\theta}.
      \label{eq:nuisance_schur}
    \end{equation}
    The corresponding metric is
    \begin{equation}
      \operatorname{CRB}_{\rm ang}
      =\operatorname{Tr}\!\left(
      \mathbf J_{\boldsymbol\theta,\rm nuis}^{-1}\right).
      \label{eq:angle_crb}
    \end{equation}
    Equations~\eqref{eq:angular_coherence}--\eqref{eq:angular_logdet} serve as low-complexity structural surrogates during candidate generation and swap refinement, whereas \eqref{eq:angle_crb} is evaluated after the inner design provides $\mathbf R_x$ and $\bm v$. Lower signature coherence can improve numerical separability, but it is not equivalent to a lower estimation CRB, which also depends on the transmit covariance, clutter whitening, scattering amplitudes, and nuisance-parameter coupling. The CRB threshold is therefore used only to screen the generated extended-target candidates before power-based selection.
        
    \subsection{Continuous Candidate Optimization and Balanced Grouping Recovery}
    \label{subsec:candidate_recovery_task}
    
    For either branch, the nonsmooth minimum bounded utility is approximated by
    \begin{equation}
      F_\tau(\overline{\mathbf v})
      =-\tau\log\!\left(
      \sum_{i\in\mathcal I}
      \exp\!\left[-\frac{\psi_i(g_i(\overline{\mathbf v}))}{\tau}\right]
      \right),
      \quad \tau>0,
      \label{eq:softmin_task}
    \end{equation}
    where $\mathcal I=\mathcal I_{\rm PT}$ or $\mathcal I_{\rm ET}$. Here, $F_\tau$ is a smooth lower approximation of the minimum and becomes tight as $\tau\to0$. With
    \begin{equation}
      \pi_i=\frac{\exp[-\psi_i(g_i)/\tau]}
      {\sum_{j\in\mathcal I}\exp[-\psi_j(g_j)/\tau]},
      \label{eq:softmin_weight_task}
    \end{equation}
    the Wirtinger gradient is
    \begin{equation}
      \nabla_{\overline{\mathbf v}^*}F_\tau
      =\sum_{i\in\mathcal I}\pi_i\psi_i'(g_i)
      \frac{\mathbf U_i\overline{\mathbf v}}{d_i},
      \label{eq:softmin_gradient_task}
    \end{equation}
    where $d_i$ is the corresponding positive normalization factor in \eqref{eq:gk_point}, \eqref{eq:gp_point}, or \eqref{eq:gt_extended}. Parameterizing $\overline{\mathbf v}=e^{j\boldsymbol\phi}$ gives
    \begin{equation}
      \nabla_{\boldsymbol\phi}F_\tau
      =2\operatorname{Im}\!\left\{
      \overline{\mathbf v}^*\odot
      \nabla_{\overline{\mathbf v}^*}F_\tau
      \right\}.
      \label{eq:phase_gradient_task}
    \end{equation}
    An Armijo line search ensures a nondecreasing $F_\tau$ for fixed $\tau$. Multiple initializations and a continuation schedule for $\tau$ are used because the unit-modulus problem remains nonconvex. Accordingly, the phase-gradient iterations provide, at most, a stationary point of each fixed-$\tau$ smoothed problem and do not guarantee a global optimum of the original nonsmooth max--min design.
    
    For a locally optimized element-wise candidate $\overline{\mathbf v}^{\star}$, the binary grouping matrix and group-wise phase vector are recovered from
    \begin{align}
      \min_{\bm G,\mathbf v}\quad
      &\|\overline{\mathbf v}^{\star}-\bm G^T\mathbf v\|^2
      \label{prob:balanced_recovery_task}\\
      \text{s.t.}\quad
      &G_{q,n}\in\{0,1\},\quad
      \sum_{q=1}^QG_{q,n}=1,\quad |v_q|=1,\nonumber\\
      &\mu_{\min}\le\sum_{n=1}^NG_{q,n}\le\mu_{\max},
      \quad q=1,\ldots,Q.
      \nonumber
    \end{align}
    The size bounds are selected around $N/Q$ and prevent one excessively large coherent group from dominating several nearly empty groups. For fixed groups, we have
    \begin{equation}
      v_q=\exp\!\left(j\arg\sum_{n\in\mathcal S_q}
      \overline v_n^{\star}\right).
      \label{eq:centroid_task}
    \end{equation}
    For fixed centroids, the capacity-constrained assignment is a minimum-cost allocation problem with costs $|\overline v_n^{\star}-v_q|^2$. With deterministic tie breaking, alternating exact assignment and centroid updates monotonically decreases \eqref{prob:balanced_recovery_task} and terminates at a coordinatewise local solution over the finite assignment set. The measure-zero case of a zero centroid sum is handled by retaining the previous phase. For $T>1$, the diversity-aware swap refinement in \eqref{eq:extended_refinement_objective} is then applied. The recovered $\mathbf v$ only initializes the fast-timescale passive-beamforming optimization; it does not replace that optimization.
    
    \subsection{Task-Dependent Candidate Screening and Selection}
    \label{subsec:task_candidate_selection}
    
    For every candidate $c\in\mathcal C$ and each of $S$ validation channel realizations drawn under the adopted S-CSI model, the fast-timescale algorithm is executed and the original constraints are rechecked after rank-one recovery. With feasibility indicator $I_s(c)$ and recovered power $P_s(c)$, define $\widehat f(c)=S^{-1}\sum_s I_s(c)$ and $\widehat P_{\rm tx}(c)=\sum_s I_s(c)P_s(c)/\sum_s I_s(c)$. The latter is evaluated only when at least one feasible recovery exists and is always reported together with $\widehat f(c)$. Using the same validation realizations, compute
    \begin{align}
      \widehat r_c(c)
      &=\frac{1}{S}\sum_{s=1}^S
      r_{\rm eff}(\mathbf H_c^{(s)}(c)),
      \label{eq:sample_comm_rank_task}\\
      \widehat\delta_c(c)
      &=\frac{1}{S}\sum_{s=1}^S
      \frac{\sigma_1^2(\mathbf H_c^{(s)}(c))}
      {\|\mathbf H_c^{(s)}(c)\|_F^2}.
      \label{eq:sample_comm_dominance_task}
    \end{align}
    Effective rank is evaluated realization by realization before averaging. In particular, $r_{\rm eff}(\mathbb E\{\mathbf H\mathbf H^H\})$ can be large even when individual instantaneous channels are nearly rank one, because averaging may mix different dominant directions. We therefore screen the empirical mean effective rank together with the dominant-mode energy fraction $\widehat\delta_c$; jointly, these metrics distinguish an energetically multi-mode channel from one whose nominal rank is sustained mainly by weak singular components.
        
    For the point-target branch, let $\mathcal C_{\rm PT}$ contain the candidates satisfying $\widehat f(c)\ge f_{\min}$, the point-target safeguards, and
    \begin{equation}
      \widehat r_c(c)\ge r_{c,\min},
      \qquad \widehat\delta_c(c)\le\delta_{c,\max}.
      \label{eq:point_screen}
    \end{equation}
    No target-subspace rank constraint is imposed because $T=1$. The selected point-target grouping is
    \begin{equation}
      c_{\rm PT}^{\star}\in
      \arg\min_{c\in\mathcal C_{\rm PT}}\widehat P_{\rm tx}(c).
      \label{prob:point_selection}
    \end{equation}
    
    For the extended-target branch, additionally compute $\widehat r_{\mathcal T}$ and $\widehat\delta_{\mathcal T}$ by replacing $\mathbf H_c$ with $\mathbf H_{\mathcal T}$ in \eqref{eq:sample_comm_rank_task}--\eqref{eq:sample_comm_dominance_task}. Define the normalized response coherence
    \begin{equation}
      \rho_{t,u}
      =\frac{|\operatorname{Tr}(\mathbf H_{gt,t}\mathbf R_x
      \mathbf H_{gt,u}^H)|}
      {\sqrt{\mathcal E_t\mathcal E_u}+\varepsilon},
      \qquad t\ne u,
      \label{eq:normalized_response_coherence}
    \end{equation}
    where $\mathcal E_t=\operatorname{Tr}(\mathbf H_{gt,t}\mathbf R_x\mathbf H_{gt,t}^H)$. Let $\widehat\rho_{\max}(c)$ and $\widehat{\operatorname{CRB}}_{\rm ang}(c)$ denote the corresponding conditional averages over feasible validation samples. The set $\mathcal C_{\rm ET}$ retains candidates satisfying $\widehat f(c)\ge f_{\min}$ and
    \begin{align}
      \widehat\rho_{\max}(c)&\le\rho_{\max},
      &\widehat{\operatorname{CRB}}_{\rm ang}(c)&\le\epsilon_\theta,
      \label{eq:extended_separation_screen}\\
      \widehat r_c(c)&\ge r_{c,\min},
      &\widehat\delta_c(c)&\le\delta_{c,\max},
      \label{eq:extended_comm_screen}\\
      \widehat r_{\mathcal T}(c)&\ge r_{\mathcal T,\min},
      &\widehat\delta_{\mathcal T}(c)&\le\delta_{\mathcal T,\max}.
      \label{eq:extended_target_screen}
    \end{align}
    The selected extended-target grouping is
    \begin{equation}
      c_{\rm ET}^{\star}\in
      \arg\min_{c\in\mathcal C_{\rm ET}}\widehat P_{\rm tx}(c).
      \label{prob:extended_selection}
    \end{equation}
    In this work, the Gaussian receive model and nuisance-parameter assumptions stated in \eqref{eq:angle_fim}--\eqref{eq:nuisance_schur} are adopted when forming $\widehat{\operatorname{CRB}}_{\rm ang}(c)$; hence, the CRB threshold in \eqref{eq:extended_separation_screen} is enforced as part of the extended-target candidate screen rather than used only for post hoc validation. If the clutter covariance becomes angle dependent or a different nuisance model is adopted, the corresponding FIM must be modified; that extension is outside the present scope.
    
    In summary, the point-target screen protects communication-channel structure, whereas the extended-target screen additionally checks desired-scatterer concentration and angular information. For $C=|\mathcal C|$ candidates and $R_g$ gradient steps, gain generation costs $\mathcal O(CR_gN(K+T))$, excluding balanced clustering and validation-stage decompositions. The extended-target branch also evaluates a $T\times T$ signature Gram matrix and the angle-FIM/CRB screen; once the required derivatives and covariance terms are formed, this includes inversion of the $T\times T$ equivalent angle-information matrix (or the corresponding nuisance-augmented Schur complement). Selection is restricted to the retained finite candidate set and the recovered solutions produced by the stated inner algorithm; no global optimality for $P_0$ is implied.

    Selecting only the candidate with the largest outer-layer utility can be misleading because high deterministic gain may coexist with poor instantaneous recovery when post-grouping channels are nearly collinear or diffuse/clutter components produce an unfavorable realization. We therefore require a minimum empirical recovery rate before ranking the retained candidates by recovered transmit power. Effective rank is used only as a structural screening metric, not as a stand-alone performance objective.
    
    \subsection{Design Interpretation and Parameter Selection}
    Table~\ref{tab:task_comparison} summarizes the task-dependent safeguards. 
    Both branches use CU gain floors and communication-rank screening. For $T=1$, no target-subspace rank constraint is meaningful because the desired response has one column. For $T>1$, however, aggregate SCNR alone can be satisfied by a dominant scatterer; per-scatterer illumination and CCB therefore complement the signature-diversity screens. Under the adopted FIM model, the angle CRB is enforced as an estimator-aware screening criterion before the surviving extended-target candidates are ranked by recovered transmit power.
    
    \begin{table}[t]
        \caption{Task-Dependent Grouping Safeguards}
        \label{tab:task_comparison}
        \centering
        \footnotesize
        \setlength{\tabcolsep}{2.5pt}
        \renewcommand{\arraystretch}{1.02}
        \begin{tabularx}{\columnwidth}{l>{\raggedright\arraybackslash}X>{\raggedright\arraybackslash}X}
            \hline
            Aspect & Point target & Extended target \\
            \hline
            Shared & CU gain floors; communication-rank screen & CU gain floors; communication-rank screen \\
            Sensing & SCNR; target illumination & SCNR; per-scatterer illumination; CCB; diversity/CRB screens \\
            Failure mode & CU-channel collinearity & CU collapse or target-response concentration \\
            \hline
        \end{tabularx}
    \end{table}

    Parameter $r$ is a normalization-dependent design weight rather than a physical power-allocation fraction and is selected by a coarse sweep. The angle-CRB threshold $\epsilon_\theta$ represents the admissible aggregate angle-estimation variance and is specified from the sensing-resolution requirement before the retained candidates are ranked by transmit power. The utility scales $\{\chi_i\}$ should span the useful gain range: overly small values saturate the bounded utility, whereas overly large values weaken the intended emphasis on the worst components. The soft-min temperature $\tau$ is decreased with warm starts. Rank thresholds satisfy $r_{c,\min}\le\min\{M,K\}$ and $r_{\mathcal T,\min}\le\min\{M,T\}$ and should account for finite-sample uncertainty; candidates close to a screen are preferably retained unless their uncertainty interval clearly violates the threshold. Group-size bounds are centered around $N/Q$ to avoid a dominant coherent group. These parameters affect candidate generation and screening only; the fast-timescale feasibility decision is made after rank-one recovery using the original QoS constraints.
    
    \section{Joint Active and Passive Beamforming Design}
    For a fixed grouping candidate, the ISAC-BS estimates the low-dimensional post-grouping cascaded CSI and alternately updates the transmit covariances and group-wise phase shifts. The resulting fixed-group problem is
    \begin{equation}
		\label{eq:38}
		\begin{array}{*{20}{l}}
			{{P_5}:}&{\mathop {\min}\limits_{\bm W_s,\bm W_c,\bm v}}{\quad}&{\operatorname{Tr}(\bm R_x)} \\
			&{\quad}{\mathrm{s.t.}}{\quad}&{\eqref{eq:13A}\text{--}\eqref{eq:13E}.}
		\end{array}
	\end{equation}

    Problem $P_5$ remains nonconvex because the sensing constraints are quartic in $\bm v$, the phase shifts have unit modulus, and the active and passive variables are coupled. Following the SDR principle in \cite{9913311}, define $\bm R_s=\bm W_s\bm W_s^H$, $\bm R_k=\bm w_k\bm w_k^H$, and $\bm V=\bm v\bm v^H$. These lifted variables satisfy $\bm R_s\succeq\mathbf0$, $\bm R_k\succeq\mathbf0$, $\bm V\succeq\mathbf0$, $V_{q,q}=1$, $\operatorname{rank}(\bm R_k)=1$, and $\operatorname{rank}(\bm V)=1$. The rank-one constraints are relaxed during block optimization and are addressed explicitly in the recovery stage.

    Since $\lambda_{\max}(\mathbf D)\le\operatorname{Tr}(\mathbf D)$ for $\mathbf D\succ\mathbf0$, we have $\mathbf D^{-1}\succeq\mathbf I_M/\operatorname{Tr}(\mathbf D)$. Hence, the exact SCNR admits the conservative lower bound
    \begin{equation}
		\label{eq:39}
		\operatorname{Tr}\left(\mathbf H_{gt}\mathbf R_x\mathbf H_{gt}^H\mathbf D^{-1}\right)
        \ge 
        \frac{\operatorname{Tr}(\mathbf H_{gt}\mathbf R_x\mathbf H_{gt}^H)}{\operatorname{Tr}(\mathbf D)}.
    \end{equation}

    For compactness, write $\mathbf H_{g,k}(\bm V)=\mathbf C_{g,k}\bm V\mathbf C_{g,k}^H$ and define $\mathbf H_{gt,t}(\bm V)$ and $\mathbf H_{gc,l}(\bm V)$ analogously to \eqref{eq:round_trip_components}. Replacing the exact SCNR constraint by the sufficient condition induced by \eqref{eq:39} and relaxing the rank constraints yields the following conservative lifted problem:
    \begin{subequations}
		\label{eq:40}
		\begin{align}
			{{P_6}:} \min_{\bm R_s,\{\bm R_k\},\bm V} &\operatorname{Tr}(\bm R_x), \nonumber\\
			\mathrm{s.t.} \quad &\left(1+\frac{1}{\Gamma_k}\right) \operatorname{Tr}(\bm R_k\mathbf H_{g,k}) \nonumber\\
            &\quad-\operatorname{Tr}(\bm R_x\mathbf H_{g,k})\ge\sigma_k^2,
            \quad k\in\mathcal K, \label{eq:40A}\\
            &\operatorname{Tr}(\mathbf H_{gt}\bm R_x\mathbf H_{gt}^H)
            -\Gamma_s\operatorname{Tr}(\mathbf H_{gc}\bm R_x\mathbf H_{gc}^H)
            \nonumber\\[-1mm]
            &\hspace{24mm}\ge M\Gamma_s\sigma_s^2, \label{eq:40B}\\
            &\operatorname{Tr}(\mathbf H_{gt,t}\bm R_x\mathbf H_{gt,t}^H)
            \ge\underline e_t,\quad t\in\mathcal T, \label{eq:40C}\\
            &\sum_{t<u}\left|\operatorname{Tr}(\mathbf H_{gt,t}\bm R_x
            \mathbf H_{gt,u}^H)\right| \le\epsilon_s, \label{eq:40D}\\
            &\bm R_s\succeq\mathbf0,\quad \bm R_k\succeq\mathbf0,
            \quad k\in\mathcal K, \label{eq:40E}\\
            &\bm V\succeq\mathbf0,\quad V_{q,q}=1,
            \quad q=1,\ldots,Q. \label{eq:40F}
		\end{align}
	\end{subequations}
    Problem $P_6$ remains nonconvex because the transmit-covariance and lifted-phase blocks are coupled. Equation~\eqref{eq:40B} is a sufficient, rather than equivalent, replacement of the exact SCNR constraint, so the resulting design is generally conservative. We optimize the two variable blocks alternately.
    
    \subsection{Optimization of Transmit Beamformers}
    For fixed $\bm V$, all channel matrices are constant and the transmit-covariance subproblem is
    \begin{equation}
		\label{eq:41}
		\begin{array}{*{20}{l}}
			{{P_7}:}&{\mathop {\min}\limits_{\bm R_s,\{\bm R_k\}}}{\quad}&{\operatorname{Tr}(\bm R_x)} \\
			&{\quad}{\mathrm{s.t.}}{\quad}&{\eqref{eq:40A}\text{--}\eqref{eq:40E}.}
		\end{array}
	\end{equation}
    For fixed $\bm V$, problem $P_7$ is a convex semidefinite program (SDP) after the absolute-value terms in the CCB are written with standard epigraph variables. Let $\{\bm R_k^{\rm opt},\bm R_s^{\rm opt}\}$ denote an optimal solution.

    \textit{Lemma 2 (Rank-one recovery for communication beamforming):}
    Let $\{\bm R_k^{\rm opt},\bm R_s^{\rm opt}\}$ be any feasible solution of $P_7$ with $\operatorname{Tr}(\bm R_k^{\rm opt}\mathbf H_{g,k})>0$. Define
    \begin{subequations}
        \label{eq:42}
        \begin{align}
            \bm w_k^\star&=\frac{\bm R_k^{\rm opt}\mathbf h_{g,k}}
            {\sqrt{\operatorname{Tr}(\bm R_k^{\rm opt}\mathbf H_{g,k})}}, \label{eq:42A}\\
            \bm R_s^\star&=\bm R_s^{\rm opt}+\sum_{k\in\mathcal K}
            \left(\bm R_k^{\rm opt}-\bm w_k^\star\bm w_k^{\star,H}\right). \label{eq:42B}
        \end{align}
    \end{subequations}
    Then $\bm R_s^\star\succeq\mathbf0$, each communication covariance $\bm w_k^\star\bm w_k^{\star,H}$ is rank one, and all constraints and the objective of $P_7$ are preserved. This follows from $\bm R_k^{\rm opt}-\bm R_k^{\rm opt}\mathbf h_{g,k}\mathbf h_{g,k}^H\bm R_k^{\rm opt}/(\mathbf h_{g,k}^H\bm R_k^{\rm opt}\mathbf h_{g,k})\succeq\mathbf0$ and from preservation of both $\bm R_x$ and the desired-signal terms. No rank-one restriction is imposed on $\bm R_s^\star$ \cite[Proposition~1]{10440056}.
    
    \subsection{Optimization of Post-grouping Phase Shifts}
    For fixed transmit covariances, the lifted phase block can be written as the following feasibility problem:
    \begin{subequations}
		\label{eq:58}
		\begin{align}
			{{P_8}:}\quad \operatorname{find}\quad&\bm V \nonumber\\
			\mathrm{s.t.}\quad
            &\left(1+\frac{1}{\Gamma_k}\right)\operatorname{Tr}(\mathbf A_k\bm V) \nonumber\\
            &\quad-\operatorname{Tr}(\mathbf B_k\bm V)\ge\sigma_k^2,
            \quad k\in\mathcal K, \label{eq:58A}\\
            &\|\bm E\|_F^2-\Gamma_s\|\bm F\|_F^2
            \ge M\Gamma_s\sigma_s^2, \label{eq:58B}\\
            &\|\bm E_t\|_F^2\ge\underline e_t,
            \quad t\in\mathcal T, \label{eq:58C}\\
            &\sum_{t<u}|\operatorname{Tr}(\bm E_t\bm E_u^H)|
            \le\epsilon_s, \label{eq:58D}\\
            &\bm V\succeq\mathbf0,\quad V_{q,q}=1,
            \quad q=1,\ldots,Q. \nonumber
		\end{align}
    \end{subequations}
    where $\mathbf A_k=\mathbf C_{g,k}^H\bm R_k\mathbf C_{g,k}$, $\mathbf B_k=\mathbf C_{g,k}^H\bm R_x\mathbf C_{g,k}$, $\bm E=\sum_{t\in\mathcal T}\bm E_t$, and $\bm F=\sum_{l\in\mathcal L}\bm F_l$, with $\bm E_t=\gamma_{s,t}\mathbf C_{gs,t}\bm V\mathbf C_{gs,t}^H\bm R_x^{1/2}$ and $\bm F_l=\gamma_{s,l}\mathbf C_{gs,l}\bm V\mathbf C_{gs,l}^H\bm R_x^{1/2}$. Constraint \eqref{eq:58A} and the positive-semidefinite (PSD) constraints are convex, whereas \eqref{eq:58B}--\eqref{eq:58D} are nonconvex in $\bm V$. Let $\bm z=\operatorname{vec}(\bm V)$ and $\bm Q_s=\bm M_E^H\bm M_E-\Gamma_s\bm M_F^H\bm M_F$. Then
    \begin{equation}
		\|\bm E\|_F^2-\Gamma_s\|\bm F\|_F^2
        =\bm z^H\bm Q_s\bm z.
        \label{eq:59}
	\end{equation}
    where
    \begin{align}
        \bm M_E=\sum_{t\in\mathcal T}\gamma_{s,t}
        \left[\left(\mathbf C_{gs,t}^H\bm R_x^{1/2}\right)^T
        \otimes\mathbf C_{gs,t}\right], \label{eq:ME}\\
        \bm M_F=\sum_{l\in\mathcal L}\gamma_{s,l}
        \left[\left(\mathbf C_{gs,l}^H\bm R_x^{1/2}\right)^T
        \otimes\mathbf C_{gs,l}\right].
        \label{eq:MF}
    \end{align}
    The matrix $\bm Q_s$ is Hermitian but generally indefinite; consequently, the quadratic sensing-feasibility condition in $P_8$ is nonconvex.

    We use successive convex approximation (SCA) to generate feasible candidates for the lifted phase block. At SCA iteration $r$, the convex desired-energy functions are replaced by their first-order global affine lower bounds:
    \begin{subequations}
		\label{eq:60}
		\begin{align}
			&\|\bm E\|_F^2 \ge 2\operatorname{Re}\{\operatorname{Tr}(\bm E^{(r)}\bm E^H)\}
            -\|\bm E^{(r)}\|_F^2, \label{eq:60A}\\
            &\|\bm E_t\|_F^2 \ge 2\operatorname{Re}\{\operatorname{Tr}(\bm E_t^{(r)}\bm E_t^H)\}
            -\|\bm E_t^{(r)}\|_F^2. \label{eq:60B}
		\end{align}
	\end{subequations}
    where $\bm E^{(r)}=\sum_{t\in\mathcal T}\bm E_t^{(r)}$ and $\bm E_t^{(r)}$ is evaluated at $\bm V^{(r)}$. These lower bounds are value- and gradient-tight at the current point. Introducing nonnegative feasibility margins yields the convex candidate-generation subproblem
    \begin{subequations}
		\label{eq:62}
		\begin{align}
			{{P_9}:} 
            \max_{\bm V,\eta_s,\{\eta_k\}} 
            &\eta_s+\sum_{k\in\mathcal K}\eta_k, \nonumber\\
			\mathrm{s.t.}\quad
            &\left(1+\frac{1}{\Gamma_k}\right)\operatorname{Tr}(\mathbf A_k\bm V) \nonumber\\
            &\quad-\operatorname{Tr}(\mathbf B_k\bm V)\ge\sigma_k^2+\eta_k,
            \quad k\in\mathcal K, \label{eq:62A}\\
            &2\operatorname{Re}\{\operatorname{Tr}(\bm E^{(r)}\bm E^H)\}
            -\|\bm E^{(r)}\|_F^2 \nonumber\\[-1mm]
            &\quad-\Gamma_s\|\bm F\|_F^2
            \ge M\Gamma_s\sigma_s^2+\eta_s, \label{eq:62B}\\
            &2\operatorname{Re}\{\operatorname{Tr}(\bm E_t^{(r)}\bm E_t^H)\}
            -\|\bm E_t^{(r)}\|_F^2 \nonumber\\[-1mm]
            &\hspace{28mm}\ge\underline e_t,
            \quad t\in\mathcal T, \label{eq:62C}\\
            &\eta_s\ge0,\quad\eta_k\ge0,\quad k\in\mathcal K, \label{eq:62D}\\
            &\bm V\succeq\mathbf0,\quad V_{q,q}=1,
            \quad q=1,\ldots,Q. \nonumber
		\end{align}
	\end{subequations}
    Problem $P_9$ is a convex quadratically constrained SDP. Because \eqref{eq:60A}--\eqref{eq:60B} are global affine lower bounds that are value- and gradient-tight at $\bm V^{(r)}$, constraints \eqref{eq:62B}--\eqref{eq:62C} constitute inner approximations of the corresponding desired-energy constraints. Feasibility of these lower-bounded constraints implies feasibility of the associated lifted energy constraints. The CCB is treated differently because its absolute bilinear cross terms do not admit the same simple inner approximation; each candidate update is therefore accepted only if the exact lifted constraints \eqref{eq:58A}--\eqref{eq:58D} are satisfied. Rank-one/unit-modulus feasibility is asserted only after Gaussian randomization and re-evaluation of the original SINR, exact SCNR, illumination, and CCB constraints. The implementation distinguishes three levels: conservative SCNR approximation, rank-relaxed lifted feasibility, and final feasibility of a physical phase vector. Because the CCB is enforced by post-update acceptance rather than a first-order-tight surrogate, stationarity of the original phase problem is not claimed.
    
    Algorithm~\ref{alg:SDR_AO} summarizes the complete update and recovery procedure.
    \begin{algorithm}
        \caption{Proposed SDR-Based AO Algorithm for Joint Active and Passive Beamforming}
        \label{alg:SDR_AO}
        \begin{algorithmic}[1]
            \REQUIRE Grouping matrix $\bm G$, post-grouping channels $\{\mathbf C_{g,k}\}$, $\{\mathbf C_{gs,t}\}$, and $\{\mathbf C_{gs,l}\}$, thresholds, and tolerances.
            \ENSURE Recovered transmit covariances $\{\bm R_s,\bm R_k\}$ and phase vector $\bm v$.
            
            \STATE Initialize $\bm V^1=\bm v^1(\bm v^1)^H$ from the grouping-recovery phase vector; use a phase-I problem or multiple starts until $P_7$ is feasible for the fixed $\bm V^1$. Set $p=1$.
            \REPEAT
                \STATE With $\bm V=\bm V^p$, solve $P_7$ to obtain $\{\bm R_s^p,\bm R_k^p\}$.
                \STATE Set $\bm V^{p,0}=\bm V^p$ and repeatedly solve $P_9$, updating its SCA point, until the phase-block change is below tolerance.
                \STATE Accept the resulting $\bm V^{p+1}$ only after verifying \eqref{eq:58A}--\eqref{eq:58D}; otherwise retain $\bm V^p$ and terminate the current start.
                \STATE Update $p = p + 1$.
            \UNTIL{The relative transmit-power decrease is below tolerance.}
            \STATE Generate unit-modulus candidates from the final $\bm V$ by Gaussian randomization. For each candidate, resolve $P_7$ and recheck the exact SINR, SCNR, illumination, and CCB constraints; retain the feasible candidate of minimum power.
            \STATE Recover $\{\bm w_k^\star\}$ using \eqref{eq:42A} and factorize $\bm R_s^\star$ to obtain the sensing streams. If no randomized candidate is feasible, declare rank-one recovery unsuccessful.
        \end{algorithmic}
    \end{algorithm}

    Each accepted phase update keeps the current transmit covariances feasible for the next $P_7$ solve; hence, the AO transmit-power sequence is nonincreasing and lower bounded, and its objective value converges. This does not imply stationarity of $P_5$, because the phase block maximizes a feasibility margin and the CCB is enforced by post-update acceptance. SDR likewise does not ensure rank-one recovery; feasibility is claimed only for a randomized physical phase that passes the original-constraint audit. The reported point is therefore a feasible finite-iteration solution, not an optimizer of $P_0$ or $P_5$.
    
    \subsection{Initialization and Numerical Safeguards}
    The grouping centroid in \eqref{eq:centroid_task} initializes $\bm v$. If $P_7$ is infeasible for this initialization, a phase-I procedure tests statistically informed rotations and additional random unit-modulus starts; only starts with nonnegative feasibility margin are retained. The AO/SCA iterations use warm starts, relative-change tolerances, and iteration caps. For numerical conditioning, solver matrices are Hermitian-symmetrized and the channels/noise powers are normalized consistently. After Gaussian randomization, every projected unit-modulus phase is followed by a new $P_7$ solve and an audit of the original constraints; candidates feasible only in the relaxed problem are discarded, and the feasible-recovery rate is reported.
    
    \subsection{Complexity Analysis}
    Problem $P_7$ contains $K+1$ Hermitian PSD blocks of size $M$, whereas $P_9$ contains one $Q\times Q$ PSD block and $K+1$ scalar margins. Let $m_7=\mathcal O(K+T^2)$, after introducing epigraph variables for the absolute-value terms, and let $m_9=\mathcal O(K+T)$ denote the corresponding numbers of scalar conic constraints. As a coarse dense-solver scaling proxy, consider the generic interior-point estimate $\mathcal C_{\rm SDP}(n,m,\epsilon) =\mathcal O\!\left(\sqrt n\log(1/\epsilon) (mn^3+m^2n^2+m^3)\right)$. A conservative order estimate for $I_{\rm AO}$ AO iterations and $I_{\rm SCA}$ SCA iterations is therefore $\mathcal O\big(I_{\rm AO}[\mathcal C_{\rm SDP}((K+1)M,m_7,\epsilon_7)+I_{\rm SCA}\mathcal C_{\rm SDP}(Q,m_9,\epsilon_9)]\big)$. Because this expression ignores exploitable block-diagonal and sparsity structure, it should be interpreted only as a generic dense-solver scaling law rather than as an implementation-level runtime prediction. Gaussian randomization adds one $P_7$ solve for each tested phase candidate.

    The phase block contains a $Q\times Q$ PSD variable rather than an $N\times N$ one. Although each slow-timescale grouping-gradient evaluation can still scale linearly with $N$, this step is amortized over many coherence intervals and does not require an $N\times N$ online SDP. Increasing $Q$ reduces grouping distortion at the cost of higher training and online optimization dimensions; Section~\ref{sec:numerical_results} evaluates this tradeoff.
    
    \section{Numerical Results}
    \label{sec:numerical_results}
    
    We evaluate five grouping strategies over $800$ independent realizations, using the same element-domain channel across strategies in each trial. A recovery is feasible only if the original SINR, exact SCNR, illumination, and CCB constraints hold within $2\times10^{-3}$ relative tolerance. Shading shows $\overline P\pm1.96s_P/\sqrt{n_f}$ over feasible powers; failed recoveries are not imputed and paired win rates use jointly feasible trials. Overall, $155{,}989$ of $156{,}000$ records are feasible, with a minimum per-curve recovery rate of $99.75\%$. Powers are arithmetic means in watts and savings are $10\log_{10}(P_{\rm ref}/P_{\rm proposed})$; a fixed master seed is used without smoothing or interpolation.
    
    \subsection{Simulation Setup and Benchmarks}

    The ISAC-BS and EG-XL-IRS are located at $(0,0)$ m and $(100,10)$ m. Two CUs are uniformly drawn from a radius-$2.5$ m disk centered at $(100,0)$ m and the path-loss-limited CU from an identical disk centered at $(115,0)$ m. The large-scale power gain is $\ell(d)=10^{-3}d^{-2.2}$. Both communication hops use a $3$-dB Rician factor. The default target has $T=3$ scattering centers separated by $15^\circ$ around broadside, with total RCS $-40$ dB divided equally among them. Two clutter scatterers lie outside $7.5^\circ$ guard intervals and each has RCS $20$ dB below the total target RCS; scattering phases are independent and uniform.
        
    For numerical conditioning, communication channels are normalized by $\sigma_k$ and one-way sensing factors by $(\sigma_s^2)^{1/4}$, so that both solver-side noise variances equal one while the recovered physical transmit power is unchanged. We set $\underline e_t=M\Gamma_s\sigma_s^2/T$ and $\epsilon_s=0.25M\Gamma_s\sigma_s^2$. The remaining default parameters are listed in Table~\ref{tab:simulation_parameters_revised}.
    
    \begin{table}[t]
    \caption{Default Simulation Parameters}
    \label{tab:simulation_parameters_revised}
    \centering
    \setlength{\tabcolsep}{3.5pt}
    \begin{tabular}{lc}
    \hline
    Parameter & Value \\
    \hline
    ISAC-BS antennas / CUs, $(M,K)$ & $(8,3)$ \\
    IRS elements / groups, $(N,Q)$ & $(4096,8)$ \\
    Target / clutter scatterers, $(T,L)$ & $(3,2)$ \\
    Rician factor / tradeoff coefficient & $3$ dB / $0.20$ \\
    CU SINR / sensing SCNR & $20$ dB / $10$ dB \\
    CU / sensing noise power & $-80$ dBm / $-80$ dBm \\
    Normalized illumination / CCB & $1$ / $0.25$ \\
    Grouping-gradient updates / swaps & $32$ / $32$ \\
    AO / SCA updates / randomizations & $3$ / $2$ / $24$ \\
    Monte Carlo realizations per point & $800$ \\
    \hline
    \end{tabular}
    \end{table}
    
    All methods use identical iteration/randomization budgets, restart rules, AO/SCA optimization, rank-one recovery, and feasibility audits, leaving grouping as the main design difference. Random grouping uses balanced partitions, adjacent-element-grouping XL-IRS (AEG-XL-IRS) uses adjacent groups, and the proposed method balances weak-CU and target S-CSI before diversity-aware swaps. Sensing- and communication-only baselines deactivate the opposite outer-layer utility ($r=0$ and $r=1$ are limiting cases). QoS sweeps keep the nominal-QoS grouping fixed with feasible warm starts, while the $Q\in\{2,4,8,16\}$ sweep uses nested partitions and coarse-to-fine phase carryover to retain inherited feasible points. Because grouping changes both coherent gain and channel geometry, we also keep restart/randomization limits and online iteration budgets identical and report recovery rates rather than allocating extra computation to difficult partitions. The curves therefore characterize the stated finite-iteration implementations, not global optima.
    
    \subsection{Communication--Sensing Weight}

    \begin{figure}[htbp]
        \centering
        \includegraphics[width=\columnwidth]{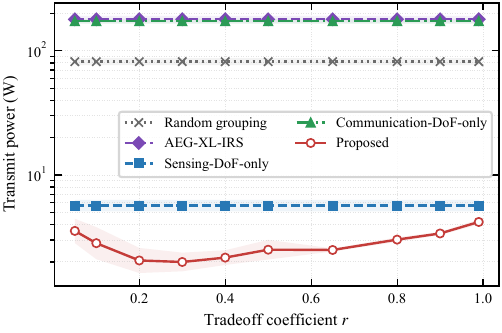}
        \caption{Average transmit power versus the communication--sensing tradeoff coefficient $r$. The four fixed baselines are repeated across $r$ as common references.}
        \label{fig:revised_dof}
    \end{figure}
    
    Fig.~\ref{fig:revised_dof} shows a broad interior minimum: the proposed method decreases from $3.552$ W at $r=0.05$ to $2.054$ W at $r=0.20$ and $1.996$ W at $r=0.30$, then rises to $4.192$ W at $r=0.99$. The overlapping intervals at $r=0.20$ and $0.30$ motivate $r=0.20$ as a sensing-aware operating point rather than a universal optimum. At $r=0.20$, Sensing-DoF-only requires $5.684$ W, giving a $63.87\%$ ($4.42$ dB) saving and a $95.75\%$ paired win rate. For this geometry and finite solver budget, the larger powers of random, adjacent, and communication-only grouping indicate that fast phase adaptation does not fully compensate for an unfavorable slow-timescale partition. A partition that overconcentrates deterministic gain in unsuitable propagation modes can reduce weak-CU separability or target-response diversity and require additional active-beamforming power to satisfy the joint SINR and sensing constraints.
        
    \subsection{Group Dimension and Pilot Overhead}

    \begin{figure}[htbp]
        \centering
        \includegraphics[width=\columnwidth]{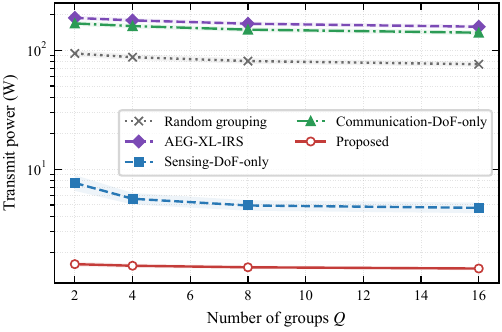}
        \caption{Average transmit power versus the number of balanced XL-IRS groups $Q$. Nested partitions and feasible-solution carryover are used for all strategies.}
        \label{fig:revised_q}
    \end{figure}
    
    With nested partitions and feasible-solution carryover, Fig.~\ref{fig:revised_q} is nonincreasing for all methods. The proposed method changes from $1.588$ W at $Q=2$ to $1.459$ W at $Q=16$, whereas Sensing-DoF-only decreases from $7.648$ to $4.724$ W, suggesting that extra online coefficients partly compensate for an unfavorable weak-CU geometry in the latter partition. At $Q=8$, the proposed method requires $1.494$ W versus $4.945$ W for Sensing-DoF-only, a $5.20$-dB saving with a $97.88\%$ paired win rate. Increasing $Q$ from $8$ to $16$ yields only $2.34\%$ further reduction while $(Q+1)/(N+1)$ rises from $0.220\%$ to $0.415\%$. For this setup, $Q=8$ is an empirical power--overhead knee rather than a universal optimum.
    
    The $Q$ sweep should not be interpreted as showing that more groups are intrinsically unnecessary. Rather, it shows diminishing returns under the present channel statistics, candidate design, and finite online optimization budget. Increasing $Q$ enlarges the passive control space but also increases the dimension of cascaded-CSI estimation and the lifted phase SDP. The preferred $Q$ depends jointly on transmit power, training overhead, online computation, and coherence time.
        
    \subsection{Communication and Sensing QoS}

    \begin{figure*}[!t]
    \centering
    \includegraphics[width=0.94\textwidth]{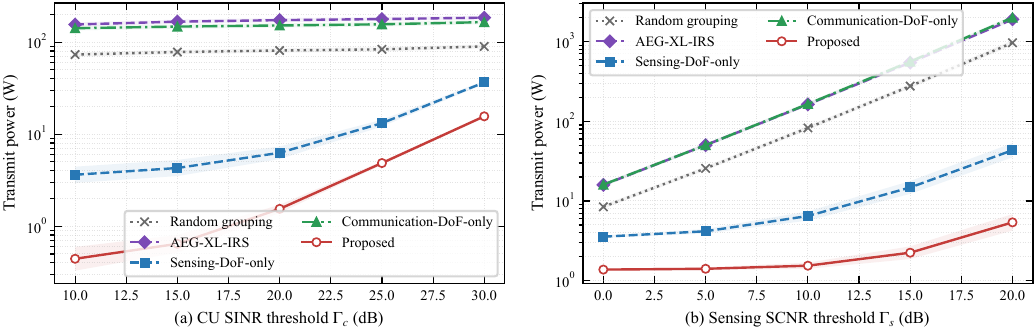}
    \caption{Average transmit power versus (a) the common CU SINR threshold and (b) the sensing SCNR threshold. The grouping matrix of each method is fixed at its nominal-QoS design during each sweep.}
    \label{fig:revised_qos}
    \end{figure*}
    
    The curves in Fig.~\ref{fig:revised_qos} are nondecreasing, consistent with nested QoS feasible sets and feasible warm starts. In Fig.~\ref{fig:revised_qos}(a), the proposed method rises from $0.443$ W at $\Gamma_c=10$ dB to $15.616$ W at $30$ dB; at the nominal $20$ dB, it saves $6.09$ dB over Sensing-DoF-only and $17.20$ dB over random grouping. At $30$ dB, the proposed method still retains a $3.68$-dB power gap over sensing-only grouping, consistent with the benefit of protecting the path-loss-limited CU. In Fig.~\ref{fig:revised_qos}(b), the proposed method increases from $1.378$ to $5.390$ W over $\Gamma_s=0$--$20$ dB, whereas Sensing-DoF-only increases from $3.563$ to $43.315$ W. At $\Gamma_s=10$ dB, the paired win rate is $97.00\%$. A sensing-favored partition can incur a large active-power penalty if it creates a deep communication fade, because the fast-timescale optimizer must then increase active-beamforming power while still meeting the sensing constraints. Two Communication-DoF-only recoveries fail at $\Gamma_c=30$ dB.
        
    \subsection{Point and Extended Targets}

    \begin{figure}[htbp]
        \centering
        \includegraphics[width=0.88\columnwidth]{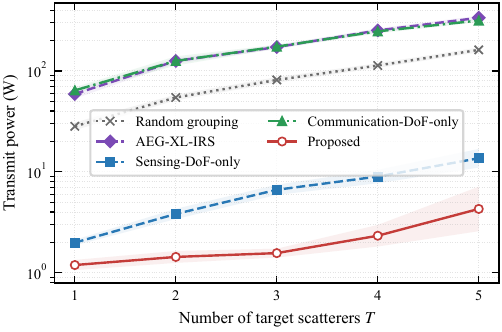}
        \caption{Average transmit power versus the number of desired target scattering centers $T$. The total target RCS is fixed and divided equally among the centers.}
        \label{fig:revised_target}
    \end{figure}
    
    In Fig.~\ref{fig:revised_target}, total target RCS is fixed while the illumination constraints grow with $T$ and pairwise CCB terms with $\binom{T}{2}$. The proposed power increases from $1.195$ W for a point target to $4.298$ W at $T=5$, reflecting the stricter coverage/decorrelation constraints rather than an increase in total target strength. At $T=3$, the saving over Sensing-DoF-only is $6.27$ dB; at $T=5$, the powers are $4.298$ and $13.648$ W, respectively, giving a $68.50\%$ reduction and a $93.00\%$ paired win rate. The larger spread at high $T$ indicates sensitivity to realizations containing both a weak CU and poorly separated target responses.
        
    \subsection{Clutter Robustness}
    
    \begin{figure}[htbp]
        \centering
        \includegraphics[width=0.88\columnwidth]{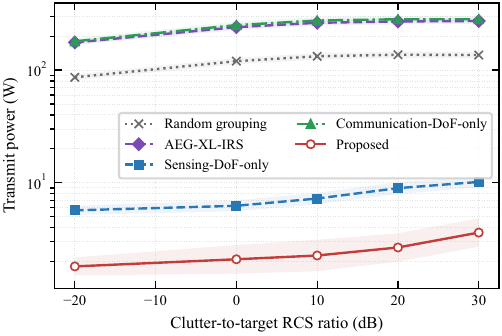}
        \caption{Average transmit power versus the RCS of each clutter scatterer relative to the total target RCS. Desired and clutter directions are independently regenerated in every realization.}
        \label{fig:revised_robustness}
    \end{figure}
    
    In Fig.~\ref{fig:revised_robustness}, increasing each clutter RCS from $-20$ to $30$ dB relative to total target RCS raises the proposed power from $1.811$ to $3.615$ W and Sensing-DoF-only from $5.692$ to $10.173$ W. At $30$ dB, the proposed method saves $4.49$ dB over sensing-only grouping and $15.75$--$18.93$ dB over the other baselines, with a $99.875\%$ feasible-recovery rate over the $10$--$30$ dB points. The gaps are consistent with the different slow-timescale partitions: baselines that preserve only one task-relevant structure can leave either weak-CU geometry or desired-target separability unfavorable, requiring the common fast-timescale optimizer to use higher transmit power. The monotonic increase with clutter strength is consistent with the SCNR denominator. Grouping can improve the geometry between desired and directional-clutter responses but cannot eliminate clutter power. Equation~\eqref{eq:diffuse_power_task} applies only to the modeled i.i.d. isotropic diffuse component; directional clutter may itself be coherently enhanced. The fast-timescale SCNR constraint handles this structured interference. Robustness to S-CSI errors, mutual coupling, hardware impairments, and unmodeled near-field scattering remains for future uncertainty-aware extensions \cite{Dong2025NearFieldRISISAC}.
    
    Across the five sweeps, the proposed method attains the smallest conditional mean power under the common online optimizer. Because all strategies use the same element-domain channel realization in each trial, the same restart/randomization budget, and the same feasibility audit, the reported gaps are reproducible consequences of the grouping partitions and their downstream recoverability under the stated implementation. Within the adopted far-field Rician model and finite solver budget, the results favor S-CSI-aware partition design; they do not establish global optimality, universal superiority, or robustness to model mismatch. All feasibility decisions use the original constraints after physical phase recovery.
        
	\section{Conclusion}
    We investigated power-efficient multiuser ISAC assisted by an element-grouping XL-IRS and developed a two-timescale design that separates slow-timescale grouping from fast-timescale active and passive beamforming. The key insight is that element grouping is not merely a means of reducing the IRS control dimension. By coherently combining phase-consistent elements, grouping can enhance desirable deterministic propagation components, but excessive concentration on a common LoS mode may simultaneously reduce the spatial diversity required for multiuser communication and extended-target sensing. This gain--rank tradeoff motivates the proposed task-adaptive grouping strategy, which exploits statistical CSI to enhance weak communication links and desired target responses while preserving task-relevant spatial dimensions. For each retained grouping pattern, the transmit covariances and group-wise reflection phases are optimized using the instantaneous post-grouping CSI, followed by physical phase recovery and verification against the original communication and sensing constraints. Numerical results demonstrate that the proposed grouping strategy consistently requires lower transmit power than the considered benchmarks under the same group dimension and online optimization budget. The results also reveal diminishing power gains as the number of groups increases, highlighting a practical tradeoff among transmit power, channel-estimation overhead, and online optimization complexity. These findings show that propagation-structure-aware element grouping can provide an effective means of scaling XL-IRS-assisted ISAC without relying on element-wise real-time control. Future work will consider statistical-CSI uncertainty, near-field propagation, hardware impairments, and joint design of grouping and channel-acquisition protocols.
     
	\bibliographystyle{IEEEtran}
	\bibliography{main}
\end{document}